\documentclass[aps,prd,preprint,groupedaddress,showpacs]{revtex4}

\usepackage{epsfig}
\usepackage{graphics}
\usepackage{slashed}
\usepackage{color}

\begin{document}
\leftmargin -2cm
\def\choosen{\atopwithdelims..}
\leftmargin -2cm
\def\choosen{\atopwithdelims..}

~~\\
 DESY~14--233 \hfill ISSN 0418-9833
\\November 2014

\title{B-meson production in the Parton Reggeization Approach \\at Tevatron and the LHC}
\unboldmath
  \author{\firstname{A.V.}
\surname{Karpishkov}} \email{karpishkov@rambler.ru}
  \affiliation{Samara State University, Ac.\ Pavlov, 1, 443011 Samara,
Russia}
\author{\firstname{V.A.}
\surname{Saleev}} \email{saleev@samsu.ru}
\affiliation{Samara State University, Ac.\ Pavlov, 1, 443011 Samara,
Russia} \affiliation{Samara State Aerospace University, Moscow
Highway, 34, 443086, Samara, Russia}
\author{\firstname{M.A. }\surname{Nefedov}}
\email{nefedovma@gmail.com}\author{\firstname{A.V.}
\surname{Shipilova}} \email{alexshipilova@samsu.ru}
\affiliation{Samara State University, Ac.\ Pavlov, 1, 443011 Samara,
Russia} \affiliation{Samara State Aerospace University, Moscow
Highway, 34, 443086, Samara, Russia}
\affiliation{{II.} Institut f\"ur Theoretische Physik, Universit\" at Hamburg,
Luruper Chaussee 149, 22761 Hamburg, Germany}

\begin{abstract}
We study the inclusive hadroproduction of $B^0$, $B^+$, and $B_s^0$ mesons
at leading order in the parton Reggeization
approach using the universal fragmentation functions extracted from the combined
$e^+e^-$ annihilation data from CERN LEP1 and SLAC SLC colliders.
We have described $B$-meson transverse momentum distributions
measured in the central region of rapidity by the CDF Collaboration
at Fermilab Tevatron and CMS Collaboration at LHC
within uncertainties and without free parameters, applying
Kimber-Martin-Ryskin unintegrated gluon distribution function in a
proton.
The forward $B$-meson production ($2.0<y<4.5$) measured by the
LHCb Collaboration also has been studied and expected disagreement
between our theoretical predictions and data has been obtained.
\end{abstract}


\maketitle

\section{Introduction}

The study of the heavy flavor production in high-energy hadronic
interactions is well suited to solve a number of tasks in particle physics.
At first, it provides a crucial test of the next-to-leading order (NLO) calculations in perturbative quantum chromodynamics (QCD)
due to the smallness of strong coupling constant $\alpha_S(\mu)$, as the lowest limit of typical
energy scale of the hard interaction $\mu$ is controlled by the bottom quark mass $m\gg\Lambda_{QCD}$, where $\Lambda_{QCD}$
is the asymptotic scale parameter of QCD.
At second, one can check the performance of different approaches to resum higher-order QCD corrections.

The experimental study was started by the first $B$-meson measurements at the CERN $S\bar p p S$ collider operating at a center-of-mass energy of $\sqrt S=0.63$~TeV~\cite{SppS}, followed in the Tevatron era by measurements of the CDF and D0 Collaborations at $\sqrt S=1.8$~TeV~\cite{Tev18} and $\sqrt S=1.96$~TeV~\cite{Tev196, Tev1961}. The quite recent results were published by the CMS
Collaboration for inclusive $B^0$- ~\cite{CMSB0},
$B^{+}$- ~\cite{CMSBplus}, and $B_s$- ~\cite{CMSBs}-meson production in proton-proton collisions at $\sqrt S=7$~TeV at the CERN Large Hadron Collider (LHC). All these measurements were implemented in the central region of rapidity, while the LHCb detector at LHC, dedicated to physics of $B$-decays, enables to measure observables of heavy flavor production in the forward rapidity region. In the Ref. ~\cite{LHCb} the LHCb team reported very recent results on $B^0$, $B^{+}$- and $B_s^0$-production in the form of transverse momentum and rapidity distributions.
Furthermore, both at Tevatron and LHC, the single and pair production of bottom-flavored jets was measured and the $b$-quark cross sections were reconstructed.
We successfully described the latter in the terms of Parton Reggeization Approach (PRA) in our previous works, see Ref.~\cite{bbTEV,bbLHC}.
But since the $b$-quark cross-section reconstruction is implemented through the observation of $B$-mesons decayed,
to consider a $B$-meson production in the framework of PRA seems to be a good test of its convenience,
completing our earlier investigations on open bottom production.

The proposal to apply the PRA in the field of heavy flavor production is caused by the fact that since the TeV-energy range is achieved,
we enter a new dynamical regime, namely the high-energy \textit{Regge limit}, characterized by the condition $\sqrt{S}\gg
\mu\gg\Lambda_{QCD}$, where the large coefficients of new type $\log^n(\sqrt S/\mu)$ appear in all-order terms of perturbative QCD series, violating its convergence.
Such a way, the new small parameter $x\sim\mu/\sqrt S$ need to be introduced and the terms proportional to $\log^n(1/x)$ should be resummed.

The small-$x$ effects cause the distinction of
the perturbative corrections relative for different processes and
different regions of phase space. At first, the higher-order
corrections for the production of heavy final states, such as Higgs
bosons, top-quark pairs, dijets with large invariant masses, or
Drell-Yan pairs, by initial-state partons with relatively large
momentum fractions $x\sim 0.1$ are dominated by soft and collinear
gluons and may increase the cross sections up to a factor 2. By
contrast, relatively light final states, such as
small-transverse-momentum heavy quarkonia, single jets, prompt
photons, or dijets with small invariant masses, are produced by the
fusion of partons with small values of $x$, typically $x\sim
10^{-3}$ because of the large values of $\sqrt{S}$. Radiative
corrections to such processes are dominated by the production of
additional hard jets. The only way to treat such processes in the
conventional collinear parton model (CPM) is to calculate
higher-order corrections in the strong coupling constant
$\alpha_S=g_S^2/4\pi$, which could be a challenging task for some
processes even at the next-to-leading order  level. To overcome
this difficulty and take into account a sizable part of the
higher-order corrections in the small-$x$ regime, the
$k_T$-factorization framework, was introduced \cite{KTCollins,KTGribov,KTCatani}.

The above mentioned $B$-meson production data at the LHC
were been previously under study in
the conventional collinear parton model of QCD
at the next-to-leading order level of accuracy in the Refs.~\cite{GMVFNB,newNLO},
and for the discussion of Tevatron data see Refs. therein.
The two working schemes were implemented: the
general-mass variable-flavor-number (GM-VFN) scheme~\cite{GMVFN},
and the so-called fixed order scheme improved with next-to-leading
logarithms (FONLL scheme)~\cite{FONLL}. In the former one, realized
in the Ref.~\cite{GMVFNB}, the large
fragmentation logarithms dominating at $p_T>>m$ are resummed through
the evolution of the nonperturbative fragmentation functions (FFs), satisfying the
Dokshitzer-Gribov-Lipatov-Altarelli-Parisi (DGLAP)~\cite{DGLAP}
evolution equations. At the same time, the full dependence on the
bottom-quark mass in the hard-scattering cross section is retained to
describe consistently $p_T\sim m$ region. The $B$-meson FFs were
extracted both at leading and next-to-leading order in the GM-VFN
scheme from the combined fit of data on $B$-meson production in $e^+e^-$ annihilation. Opposite, in the FONLL approach, the NLO
$B$-meson production cross sections are calculated with a
nonperturbative $b$-quark FF, that is not a subject to
DGLAP~\cite{DGLAP} evolution. The FONLL scheme was implemented in
the Refs.~\cite{newNLO} and its main ingredients are the
following: the NLO fixed order calculation (FO) with resummation of
large transverse momentum logarithms at the next-to-leading level
(NLL) for heavy quark production. For the consistency of the
calculation, the NLL formalism should be used to extract the
nonperturbative FFs from $e^+e^-$ data, and in the
Refs.~\cite{newNLO} the scheme of calculation of heavy quark
cross section and extraction
 of the nonperturbative FFs are directly connected and must be used only together.
In general, the theoretical predictions obtained in
Refs.~\cite{newNLO,GMVFNB} describe data well within uncertainties.

 The first study of open beauty hadroproduction in the alternative high-energy factorization scheme, namely the $k_T$-factorization framework~\cite{KTCollins,KTGribov,KTCatani},
 was firstly performed
  in the Ref.~\cite{TeryaevBB},
  where a good description of the $b\bar b$-pair production data from Tevatron, Run I, was acquired.
The authors operated with off-shell initial gluons and the formalism of transverse-momentum
  dependent parton distributions. In the present work we develop this approach, introducing the $k_T$-factorization framework together with the fully gauge-invariant amplitudes with \textit{Reggeized} gluons in the initial state. This combination we call the Parton Reggeization Approach everywhere below. We suppose PRA to be more theoretically reasonable than previous studies in $k_T$-factorization,
  as it is based on a gauge invariant effective theory for the QCD processes which occur in the quasi-multi-Regge kinematics.
   Therefore it preserves the gauge invariance of high-energy particle
  production amplitudes and allows a consistent continuation towards the NLO calculations.

Recently, PRA was successfully applied to analyze as the processes which involve heavy quark production:
bottom-flavored jets
\cite{bbTEV,bbLHC}, charmonium and bottomonium production
\cite{KniehlSaleevVasin1,KniehlSaleevVasin2,PRD2003,NSS_charm,NSS_bot}, as
the number of others: inclusive
production of single jet \cite{KSSY}, pair of jets \cite{NSSjets},
prompt-photon \cite{tevatronY,heraY}, photon plus jet \cite{KNS14},
Drell-Yan lepton pairs \cite{NNS_DY},
at Tevatron and the LHC. These studies  have demonstrated the
advantages of the high-energy factorization scheme used in PRA for
the description of data, compared with the calculations in collinear parton model.

This paper is organized as follows.
In Sec.~\ref{sec:two} we present basic formalism of our
calculations, the PRA and the fragmentation model. In
Sec.~\ref{sec:three} our results are presented in comparison with the
experimental data and discussed. In Sec.~\ref{sec:four} we summarize our conclusions.

\section{Basic Formalism}
\label{sec:two}

We study the production of $B$-mesons with high transverse momenta much larger than a $b$-quark mass.
In this region we can apply the so-called massless scheme or zero-mass variable-flavor-number scheme (ZM-VFNS)~\cite{BKK1998,Cacciari1994}
treating a $b$-quark as a massless parton.
For this case the $B$-cross section
can be written in a factorized form as it stated by
the factorization theorem of QCD~\cite{MeleNason}:
\begin{eqnarray}
\frac{d\sigma(p+p\to B+ X)}{dp_{BT} dy}= \sum_i \int_0^1
\frac{dz}{z} D_{i\to B}(z,\mu^2) \frac{d\sigma(p+p\to i(p_i)+
X)}{dp_{iT}dy_i},\label{eq:frag}
\end{eqnarray}
where $D_{i\to B}(z,\mu^2)$ is the fragmentation function for producing the $B$-meson
from the parton $i$, created at the hard scale $\mu$, the fragmentation parameter $z$ is defined through the relation $p_i=p_B/z$,
with $p_B$ and $p_i$ to be $B$-meson and
$i$-parton four-momenta, correspondingly, and their rapidities $y_B=y_i$.
The high-transverse-momenta $b$-quark radiates a large amount of its energy in the form of hard, collinear gluons,
causing the presence of the logarithms of the form $\alpha_S\log(\mu^2/m_b^2)$ in all orders of perturbative series.
These large logarithms can be resummed through the Dokshitzer-Gribov-Lipatov-Altarelli-
Parisi (DGLAP) evolution equations for nonperturbative fragmentation functions (FFs).
The latter can be obtained only from experiment.
In the Ref.~\cite{FFB}, the nonperturbative FFs for the transitions $a\to B$, where $a$ is any parton, including $b$ and $\bar b$ quarks, were extracted at NLO in the $\overline{MS}$ factorization scheme with $n_f = 5$ flavors from the experimental data for the reaction $e^+e^-\to B+X$ provided by the ALEPH~\cite{ALEPH} and OPAL~\cite{OPAL} Collaborations at the CERN LEP1 collider and by the SLD Collaboration~\cite{SLD} at the SLAC SLC collider. These data were taken on the $Z$-boson peak, that strongly suppresses the finite-$m_b$ effects which are of relative order $m^2_b/m^2_Z$, giving the internal consistence of resulting FFs with the ZM-VFN scheme which we keep throughout our analysis.
As input, in the fits of Refs.~\cite{FFB}, the
parameterizations at the initial scale $\mu_0=m_b$ for the FF's were
taken in the simple power ansatz.

\begin{figure}[ht]
\begin{center}
\includegraphics[width=0.8\textwidth, angle=-90, clip=]{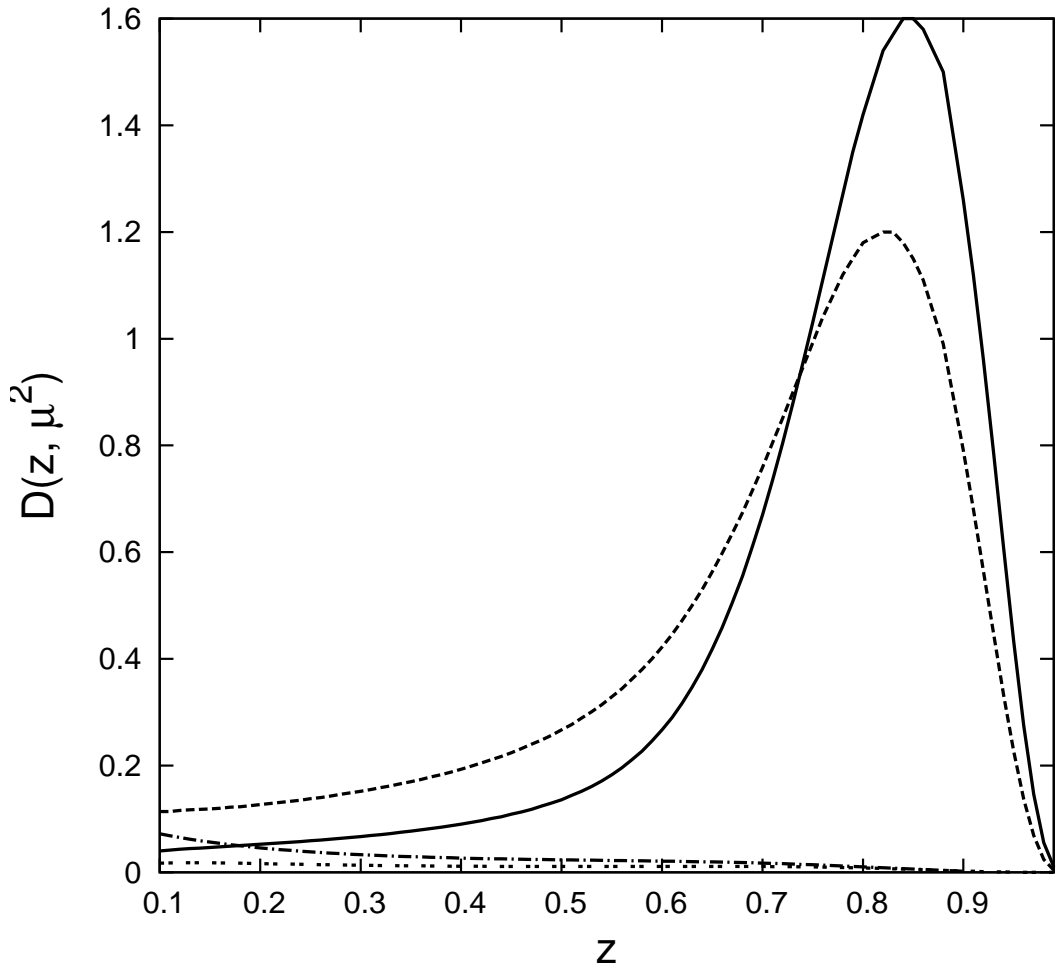}
\end{center}
\caption{The fragmentation function $D(z,\mu^2)$ of $b$-quarks and gluons into $B$ mesons from Ref.~\cite{FFB} at the
 $\mu^2=100$ GeV$^2$ (solid curve for $b$-quark, pair-dotted for gluon) and
 $\mu^2=1000$ GeV$^2$ (dashed line for $b$-quark, dash-dotted for gluon).
 \label{fig:FFB}}
\end{figure}

It was shown in Ref.~\cite{FFB}, that the major part of $B$-mesons is produced through the
gluon and bottom quark fragmentation, while the light
quark fragmentation turns out to be negligible. Following this, in our
study we will consider the $b$-quark and gluon fragmentation into different
$B$-mesons only. To illustrate a difference of contributions to the $B$-meson production we show in the Fig.~\ref{fig:FFB} the $b-$quark and gluon FF's into $B$-meson.

At high energies the bottom quarks are dominantly created via direct parton-parton collisions.
When the center-of-mass energy is much larger than the bottom quark mass,
the prior role is played by the gluon-gluon fusion.
In hadron collisions the cross sections of processes with a hard
scale $\mu$ can be represented as a convolution of scale-dependent
parton (quark or gluon) distributions and squared hard parton
scattering amplitude. These distributions correspond to the density
of partons in the proton with longitudinal momentum fraction $x$
integrated over transverse momentum up to $k_T=\mu$. Their evolution
from some scale $\mu_0$, which controls a non-perturbative regime,
to the typical scale $\mu$ is described by DGLAP~\cite{DGLAP}
evolution equations which allow to sum large logarithms of type
$\log(\mu^2/\Lambda_{QCD}^2)$ (collinear logarithms). The typical
scale $\mu$ of the hard-scattering processes is usually of order of
the transverse mass $m_T=\sqrt{m^2+|{\bf p}_T|^2}$ of the produced
particle (or hadron jet) with (invariant) mass $m$ and transverse
two-momentum ${\bf p}_T$. With increasing energy, when the ratio of
$x \sim \mu/\sqrt S$ becomes small, the new large logarithms
$\log(1/x)$, soft logarithms, are to appear and can become even more
important than the collinear ones. These logarithms present both in
parton distributions and in partonic cross sections and can be
resummed by the Balitsky-Fadin-Kuraev-Lipatov (BFKL)
approach~\cite{BFKL}.  The approach gives the description of QCD
scattering amplitudes in the region of large $S$ and fixed momentum
transfer $t$, $S \gg |t|$ (Regge region), with various color states
in the $t$-channel.
 Entering this region requires us to reduce approximations to keep the true kinematics of the process.
 It becomes possible introducing the unintegrated over transverse momenta parton distribution functions
 (UPDFs) $\Phi(x,t,\mu^2)$, which depend on parton transverse momentum
${\bf q}_T$ while its virtuality is $t=-|{\bf q}_T|^2$.  The UPDFs
are defined to be related with collinear ones through the equation:
\begin{eqnarray}
xG(x,\mu^2)=\int^{\mu^2}dt \Phi(x,t,\mu^2).
\end{eqnarray}
The UPDFs satisfy the BFKL evolution
equation \cite{BFKL} which is suited to resum soft logarithms and appear in the BFKL approach as a particular result in
the study of analytical properties of the forward scattering amplitude.
The basis of the BFKL approach is the gluon Reggeization \cite{gRegge}, as at small $x$ the gluons are the dominant partons.

The gluon Reggeization appears considering special types of
kinematics of processes at high-energies. At large $\sqrt S$ the
dominant contributions to cross sections of QCD processes gives
multi-Regge kinematics (MRK). MRK is the kinematics where all
particles have limited (not growing with $\sqrt S$) transverse
momenta and are combined into jets with limited invariant mass of
each jet and large (growing with $\sqrt S$) invariant masses of any
pair of the jets. At leading logarithmic approximation of the BFKL
approach (LLA), where the logarithms of type $(\alpha_s\log(1/x))^n$
are resummed, only gluons can be produced and each jet is actually a
gluon. At next-to-leading logarithmic approximation (NLA) the terms
of $\alpha_s(\alpha_s\log(1/x))^n$
 are collected and a jet can contain a couple of partons (two
gluons or quark-antiquark pair). Such kinematics is called quasi
multi-Regge kinematics. Despite of a great number of
contributing Feynman diagrams it turns out that at the Born level in
the MRK amplitudes acquire a simple factorized form. Moreover,
radiative corrections to these amplitudes do not destroy this form,
and their energy dependence is given by Regge factors
$s_i^{\omega(q_i)}$, where $s_i$ are invariant masses of couples of
neighboring jets and $\omega(q_i)$ can be interpreted as a shift of
gluon spin from unity, dependent from momentum transfer $q$. This
phenomenon is called gluon Reggeization.

Due to the Reggeization of quarks and gluons, an important role is
dedicated to the vertices of Reggeon-particle interactions. In
particular, these vertices are necessary for the determination of
the BFKL kernel. To define them we can notice the two ways: the
"classical" BFKL method~\cite{FadinFiore} is based on analyticity
and unitarity of particle production amplitudes and  the properties
of the integrals corresponding to the Feynman diagrams with two
particles in the $t$-channel has been developed. Alternatively, they
can be straightforwardly derived from the non-Abelian
gauge-invariant effective action for the interactions of the
Reggeized partons with the usual QCD partons, which was firstly
introduced in Ref.~\cite{KTLipatov} for Reggeized gluons only, and
then extended by inclusion of Reggeized quark fields in the
Ref.~\cite{LipVyaz}. The full set of the induced and effective
vertices together with Feynman rules one can find in
Refs.~\cite{LipVyaz,KTAntonov}.

Recently, an alternative method to obtain the gauge-invariant $2\to n$ amplitudes with off-shell initial-state partons, which is mathematically equivalent to the PRA, was proposed in Ref.~\cite{Kutak}. These $2\to n$ amplitudes are extracted by using the spinor-helicity representation with complex momenta from the auxiliary $2\to n+2$ scattering processes which are constructed to include the $2 \to n$ scattering processes under consideration. This method is more suitable for the implementation in automatic matrix-element generators, but for our study the use of Reggeized quarks and gluons is found to be simpler.

As we mentioned above, we will consider the $B$-meson production by only the $b$-quark and gluon fragmentation.
The lowest order in $\alpha_S$ parton subprocesses of PRA in which gluon or $b$-quark are produced are the following:
a gluon production
via two Reggeized gluon fusion
\begin{eqnarray}
\mathcal {R} + \mathcal {R} \to g,\label{eq:RRg}
\end{eqnarray}
and the corresponding quark-antiquark pair production
\begin{eqnarray}
\mathcal {R} + \mathcal {R} \to b + \bar b, \label{eq:RRQQ}
\end{eqnarray}
where $\mathcal {R}$ are the Reggeized gluons.

\begin{figure}[h]
\begin{center}
\includegraphics[width=.9\textwidth, clip=]{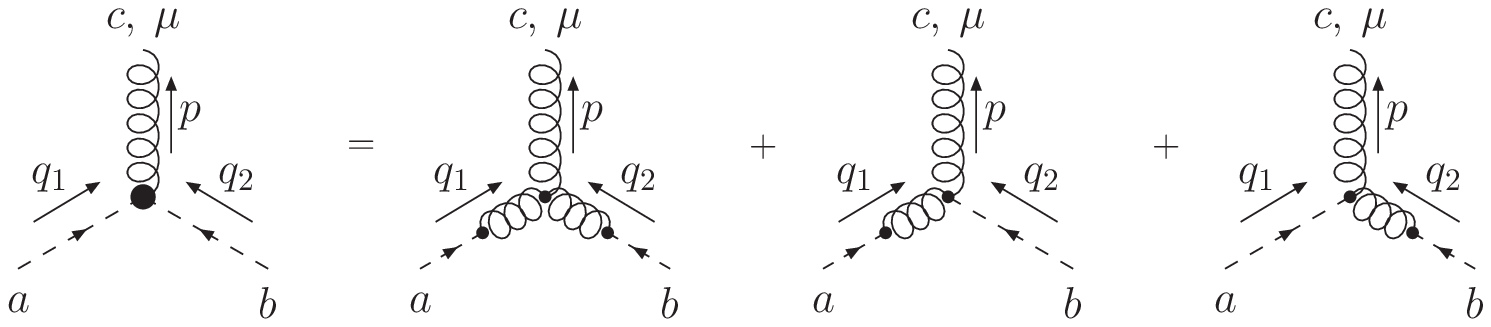}
\caption{Feynman diagrams for the subprocess (\ref{eq:RRg}). \label{fig:RRg}}
\end{center}
\end{figure}

\begin{figure}[h]
\begin{center}
\includegraphics[width=.9\textwidth, clip=]{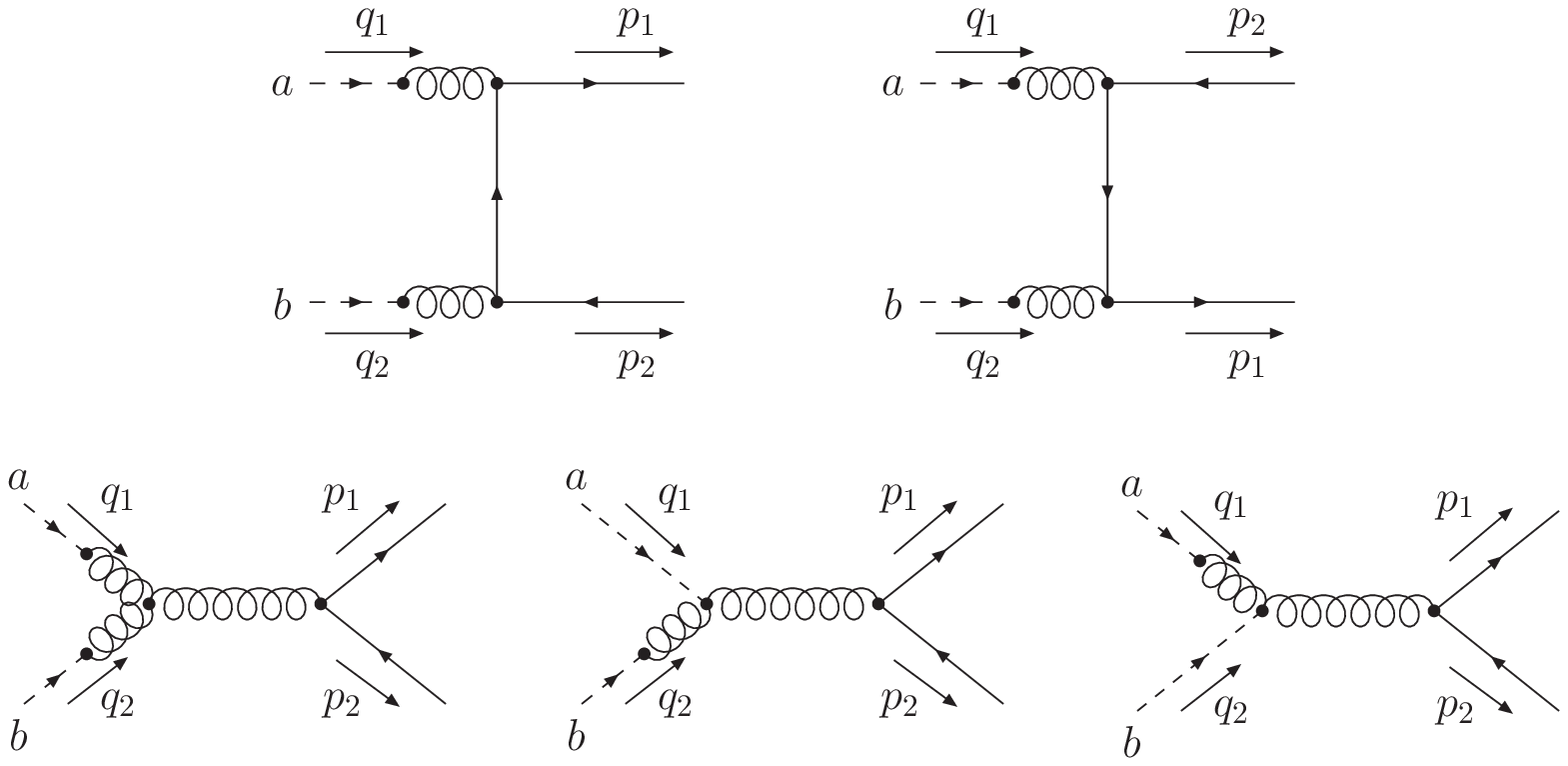}
\caption{Feynman diagrams for the subprocess (\ref{eq:RRQQ}).
\label{fig:RRQQ}}
\end{center}
\end{figure}

According to the prescription of Ref.~\cite{KTAntonov}, the
amplitudes of relevant processes (\ref{eq:RRg}) and (\ref{eq:RRQQ})
can be obtained from the Feynman diagrams depicted in
Figs.~\ref{fig:RRg} and \ref{fig:RRQQ}, where the dashed lines
represent the Reggeized gluons. Of course, the last three Feynman
diagrams in Fig.~\ref{fig:RRQQ} can be combined into the effective
particle-Reggeon-Reggeon  (PRR) vertex~\cite{KTAntonov}.

Let us define four-vectors $(n^-)^\mu=P_1^\mu/E_1$ and
 $(n^+)^\mu=P_2^\mu/E_2$, where $P_{1,2}^\mu$ are the four-momenta of
the colliding protons, and $E_{1,2}$ are their energies. We have
$(n^\pm)^2=0$, $n^+\cdot n^-=2$, and $S=(P_1+P_2)^2=4E_1E_2$. For
any four-momentum $k^\mu$, we define $k^\pm=k\cdot n^\pm$.  The
four-momenta of the Reggeized gluons can be represented as
\begin{eqnarray}
&&q_1^\mu = \frac{q_1^+}{2}(n^-)^\mu+q_{1T}^\mu\mbox{,}\nonumber\\
&&q_2^\mu = \frac{q_2^-}{2}(n^+)^\mu+q_{2T}^\mu\mbox{,}
\end{eqnarray}
where $q_{T}=(0,{\bf q}_{T},0)$
The amplitude of gluon production in fusion of two Reggeized gluons
can be presented as scalar product of Fadin-Kuraev-Lipatov effective
PRR vertex $C_{\mathcal{RR}}^{g,\mu}(q_1,q_2)$ and polarization
four-vector of final gluon $\varepsilon_\mu(p)$:
\begin{equation}
{\cal M}(\mathcal{R}+\mathcal{R}\to
g)=C_{\mathcal{RR}}^{g,\mu}(q_1,q_2)\varepsilon_\mu(p),
\end{equation}
where
\begin{eqnarray}
C_{\mathcal{RR}}^{g,\mu}(q_1,q_2)&=&-\sqrt{4\pi\alpha_s}f^{abc}
\frac{q_1^+q_2^-}{2\sqrt{t_1t_2}} \left[\left(q_1-q_2\right)^\mu+
\frac{(n^+)^\mu}{q_1^+}\left(q_2^2+q_1^+q_2^- \right)\right.\nonumber\\
&-&\left.\frac{(n^-)^\mu}{q_2^-}\left(q_1^2+q_1^+q_2^-\right)\right],
\label{amp:RRg}
\end{eqnarray}
$a$ and $b$ are the color indices of the Reggeized gluons with
incoming four-momenta $q_1$ and $q_2$, and $f^{abc}$ with
$a=1,...,N_c^2-1$ is the antisymmetric structure constants of color
gauge group $SU_C(3)$. The squared amplitude of the partonic
subprocess $\mathcal{R}+\mathcal{R}\to g$ is straightforwardly found
from Eq.~(\ref{amp:RRg}) to be
\begin{equation}
\overline{|{\cal M}(\mathcal{R}+\mathcal{R}\to g)|^2}=\frac{3}{2}\pi
\alpha_s \mathbf{p}_T^2. \label{sqamp:RRg}
\end{equation}

The amplitude of the process (\ref{eq:RRQQ}) can be presented in a
same way, as a sum of three terms ${\cal
M}(\mathcal{R}+\mathcal{R}\to b+\bar b)={\mathcal M}_1+{\mathcal
M}_2+{\mathcal M}_3$:
\begin{eqnarray}
{\mathcal M}_1&=&-i  \pi \alpha_s \frac{q_1^+ q_2^-}{\sqrt{t_1 t_2}}
T^aT^b \bar U(p_1) \gamma^\alpha\frac{\hat p_1-\hat
q_1}{(p_1-q_1)^2}\gamma^\beta V(p_2)(n^+)^\alpha(n^-)^\beta,
\nonumber\\
{\mathcal M}_2&=&-i \pi \alpha_s \frac{q_1^+ q_2^-}{\sqrt{t_1 t_2}}
T^bT^a\bar U(p_1) \gamma^\beta\frac{\hat p_1-\hat
q_2}{(p_1-q_2)^2}\gamma^\alpha
V(p_2)(n^+)^\alpha(n^-)^\beta,\\  \label{eq:ampRRQQ} {\mathcal
M}_3&=&2 \pi \alpha_s \frac{q_1^+ q_2^-}{\sqrt{t_1 t_2}}
T^cf^{abc}\frac{\bar U(p_1)\gamma^\mu V(p_2)}{(p_1+p_2)^2}
[(q_1-q_2)^\mu+\\\nonumber
&&(n^-)^\mu(q_2^++\frac{q_2^2}{q_1^-})-(n^+)^\mu(q_1^-+\frac{q_1^2}{q_2^+})],
\nonumber
\end{eqnarray}
where $T^a$ are the generators of the fundamental representation of
the color gauge group $SU_C(3)$.

The squared amplitudes can be presented as follows
\begin{eqnarray}
&&\overline{|{\mathcal M}(\mathcal{R}+\mathcal{R} \to b + \bar b)|^2}
= 256 \pi^2 \alpha_s^2 \left( \frac{1}{2 N_c} {\cal A}_{\mathrm{Ab}}
+ \frac{N_c}{2 (N_c^2 - 1)} {\cal A}_{\mathrm{NAb}} \right) \label{sqamp:RRQQ}
\end{eqnarray}

\begin{eqnarray}
&&{\cal A}_{\mathrm{Ab}} = \frac{t_1 t_2}{{\hat t} {\hat u}} -
\left( 1 + \frac{p_2^+}{{\hat
u}}(q_1^--p_2^-)+\frac{p_2^-}{{\hat t}}(q_2^+-p_2^+) \right)^2
\end{eqnarray}

\begin{eqnarray}
{\cal A}_{\mathrm{NAb}} &=& \frac{2}{S^2}\left(\frac{p_2^+
(q_1^--p_2^-)S}{{\hat u}}+\frac{S}{2}+\frac{\Delta}{\hat
s}\right)\left(\frac{p_2^- (q_2^+-p_2^+)S}{{\hat
t}}+\frac{S}{2}-\frac{\Delta}{\hat s}\right)\nonumber\\
&&-\frac{t_1 t_2}{q_1^- q_2^+ {\hat s}}\left(\left(\frac{1}{{\hat
t}}-\frac{1}{{\hat u}}\right)(q_1^- p_2^+ - q_2^+
p_2^-)+\frac{q_1^- q_2^+ {\hat s}}{{\hat t} {\hat u}}-2\right)
\end{eqnarray}

\begin{eqnarray}
&&\Delta = \frac{S}{2}\left({\hat u} - {\hat t}+2 q_1^- p_2^+-2
q_2^+ p_2^- +t_1 \frac{q_2^+-2 p_2^+}{q_2^+}  -t_2 \frac{q_1^--2
p_2^-}{q_1^-} \right)
\end{eqnarray}
Here the bar indicates averaging (summation) over initial-state
(final-state) spins and colors, $t_1 = - q_1^2 = |{\bf q}_{1T}|^2$,
$t_2 = - q_2^2 = |{\bf q}_{2T}|^2$, and
\begin{eqnarray}
&&\hat s = (q_1 + q_2)^2 = (p_1 + p_2)^2\mbox{,}\nonumber\\
&&\hat t = (q_1 - p_1)^2 = (q_2 - p_2)^2\mbox{,}\nonumber\\
&&\hat u = (q_2 - p_1)^2 = (q_1 - p_2)^2\nonumber
\mbox{.}
\end{eqnarray}
The squared amplitude (\ref{sqamp:RRQQ}) analytically coincide with
the previously obtained in Ref.~\cite{KTCollins}. We checked that in
the collinear limit, i.e. $q_{(1,2)T}\to 0$, the squared amplitude
(\ref{sqamp:RRQQ}) after averaging over the azimuthal angles
transforms to the squared amplitude of the corresponding parton
subprocess in collinear model, namely  $g+g\to b+\bar b$. We perform
our analysis in the region of $\sqrt S, p_T\gg m_b$, that allows us
to use zero-mass
 variable-flavor-number-scheme (ZM VFNS), where the masses of the charm  quarks in the hard-scattering amplitude are neglected.

In the $k_T$-factorization, differential cross section for the $2\to
1$ subprocess (\ref{eq:RRg}) has the form:
\begin{eqnarray}
\frac{d \sigma}{dy dp_T}(p + p \to g + X)=  \frac{1}{p_T^3} \int
d\phi_1 \int dt_1 \Phi(x_1,t_1,\mu^2) \Phi(x_2,t_2,\mu^2)\times\\\nonumber
\overline{|{\cal M}(\mathcal{R} + \mathcal{R} \to g)|^2} \mbox{,}
\label{eq:QMRKg}
\end{eqnarray}
where $\phi_1$ is the azimuthal angle between ${\bf p}_T$ and ${\bf
q}_{1T}$.

Analogous formula for the $2\to 2$ subprocess (\ref{eq:RRQQ}) can be
written as
\begin{eqnarray}
\frac{d\sigma}{dy_1dy_2dp_{1T}dp_{2T}}(p + p \to b(p_1)+\bar
b(p_2) + X)= \frac{p_{1T}p_{2T}}{16 \pi^3} \int d\Delta\phi \times\\\nonumber \int d\phi_1
\int dt_1 \Phi(x_1,t_1,\mu^2) \Phi(x_2,t_2,\mu^2)
\frac{\overline{|{\cal M}(\mathcal{R} + \mathcal{R} \to c + \bar
c)|^2}}{(x_1x_2 S)^2}\mbox{,} \label{eq:QMRKqq}
\end{eqnarray}
where $x_1=q_1^+/P_1^+$, $x_2=q_2^-/P_2^-$, $\Delta\phi$ is the
azimuthal angle between ${\bf p}_{1T}$ and ${\bf p}_{2T}$, the
rapidity of the final-state parton with four-momentum $p$
 is $\displaystyle{y=\frac{1}{2}\ln (\frac{p^+}{p^-})}$.
Again, we have checked a fact that in the limit of $t_{1,2}\to 0$,
we recover the conventional factorization formula of the collinear
parton model from (\ref {eq:QMRKg}) and (\ref{eq:QMRKqq}).

The important ingredient of the our scheme is unintegrated gluon distribution function, which we take as one proposed by Kimber,
  Martin and Ryskin (KMR) \cite{KMR}. These distributions are obtained introducing a single-scale auxiliary function which
  satisfies the unified BFKL/DGLAP evolution equation, where the leading BFKL logarithms $\alpha_S\log (1/x)$ are fully resummed
  and even a major (kinematical) part of the subleading BFKL effects are taken into account.
 This procedure to obtain UPDFs requires less computational efforts than
  the precise solution of two-scale evolution equations such as, for instance, Ciafaloni-Catani-Fiorani-Marchesini
  equation~\cite{CCFM}, but we found it to be suitable and adequate
  to physics of processes under study.

The usage of the $k_T$-factorization formula and UPDFs with one
longitudinal (light-cone) kinematic variable ($x$) requires the
Reggeization of the $t-$channel partons. Accordingly to Refs.~\cite{KTLipatov,LipVyaz},
Reggeized partons carry only one large light-cone component
of the four-momentum and, therefore, it's virtuality is dominated by
the transverse momentum. Such kinematics of the $t-$channel partons
corresponds to the MRK of the initial state radiation and particles,
produced in the hard process. In our previous
analysis~\cite{KniehlSaleevVasin1,KniehlSaleevVasin2,PRD2003,NSS_charm,NSS_bot}
devoted to the similar processes of heavy
  meson production we proved that these UPDFs give the best description of
the heavy quarkonium  $p_T-$spectra measured  at Tevatron
\cite{CDF} and the LHC \cite{LHCdataQuarkonium}.

As the contribution of gluon fragmentation at $\mu>\mu_0$ is initiated by the perturbative transition of gluons
to $b\bar b$-pairs encountered by DGLAP evolution equations, the part of $b$-quarks produced in the subprocess (\ref{eq:RRQQ})
with their subsequent transition to $B$-mesons are already taken into account considering $B$-meson production via gluon fragmentation.
 The simplest way to avoid double counting is to effectively subtract this contribution
 by the imposing of the lower  cut on $\hat s$ at the threshold of
 the production of the $b\bar b$ pair in (\ref{eq:QMRKqq}), i.e
 $\hat s>4m_b^2$. The precise study of
  double-counting terms and other finite-mass effects needs a separate consideration and can be a subject of our future works.

\section{Results}
\label{sec:three}
\begin{figure}[ht]
\begin{center}
\includegraphics[width=0.9\textwidth]{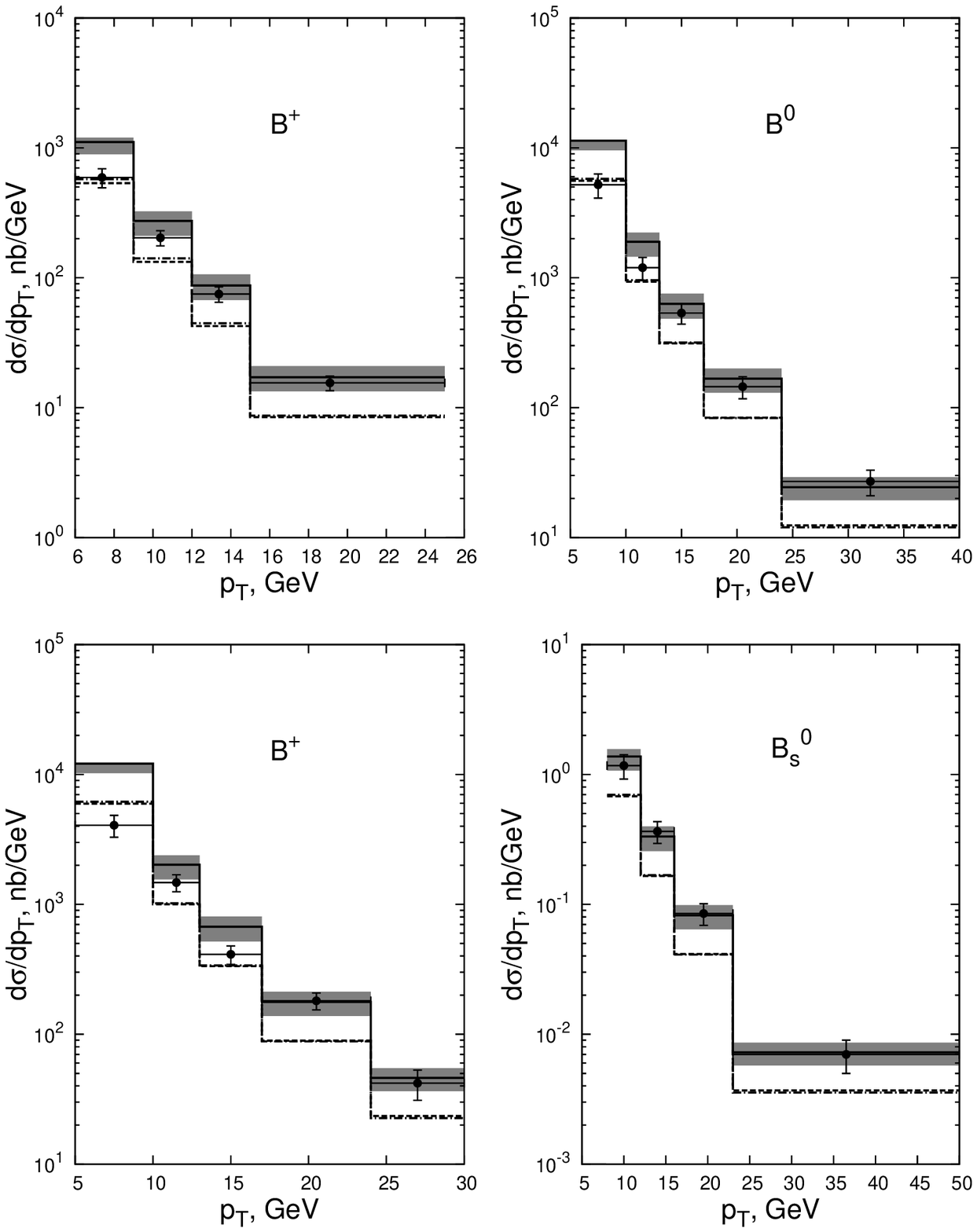}
\end{center}
\caption{Transverse momentum distributions of $B^+$-meson production at Tevatron, $\sqrt S=1.96$~TeV (left-top); $B^0$ (right-top), $B^{+}$ (left-bottom), and $B_s^0$ (right-bottom) mesons at LHC, $\sqrt S=7$~TeV.
Dashed line represents the contribution of gluon fragmentation,
dash-dotted line -- the $b$-quark-fragmentation contribution, solid line is their sum.
The CDF data at Tevatron are from the Ref.~\cite{Tev196}, the CMS data at LHC are from the Refs.~\cite{CMSB0,CMSBplus,CMSBs}, correspondingly.\label{fig:central}}
\end{figure}

We consequently come to the comparison of our predictions for the cross section distributions with experimental data.
To illustrate the rise of the signals of high-energy-asymptotic effects due to increasing of the collision energy,
we start our analysis from the data collected for the $B^+$-mesons at the collision energy of $\sqrt S=1.96$~TeV by the CDF Collaboration at Fermilab Tevatron, Run II~\cite{Tev196}.
The $B^+$-mesons were produced in the central region of rapidity
$|y|<1.0$ carrying transverse momenta up to 25~GeV. In the
Fig.~\ref{fig:central}, left-top panel, we introduce these data coming as
differential cross sections $d\sigma/dp_T$, where the particle and
antiparticle contributions are averaged, in comparison with our
predictions in the LO of the PRA. The dashed lines represent contributions of
the process (\ref{eq:RRg}) while dash-dotted lines correspond to
ones of the process (\ref{eq:RRQQ}). The sum of both contributions
is shown as a solid line. A theoretical uncertainty
is estimated by varying factorization and
renormalization scales between $1/2 m_T$ and $2 m_T$
around their central value of $m_T$, the transverse mass of a
fragmenting parton. The resulting uncertainty is depicted in the
figures by shaded bands.
We follow our comparison increasing the collision energy
but staying within the central rapidity region,
turning to the description of
the recent data from the LHC at $\sqrt S=7$~TeV collected by the CMS
Collaboration for $B^0$ mesons at $|y|<2.2$~\cite{CMSB0}, $B^+$ and $B_s^0$ mesons at $|y|<2.4$~\cite{CMSBplus,CMSBs}. In the Fig.~\ref{fig:central},
right-top, left-bottom and right-bottom panels, we show the $p_T$-distributions for $B^0$, $B^+$, and $B_s^+$ mesons, correspondingly.
At both collision energies considered we find a good agreement between our
predictions and experimental data for the large values of
$B$-meson transverse momenta, within experimental
 and theoretical uncertainties, while in
the lower $p_T$ range our predictions are found to overshoot the data, except the $B_s^+$ meson case,
where a nice coincidence for all values of $p_T$ is obtained.
But since we neglect the $b$-quark mass, the predictions in the region $p_T\sim m_b$
are obviously expected to overestimate the data and therefore should not confuse the reader.
Comparing with the previous investigations at the NLO level of CPM,
our results obtained at the LO of PRA nearly coincide with the recent ones derived in the framework of GM-VFN scheme~\cite{FFB,GMVFNB}. Considering the relative contributions of the subprocesses, in general we find the MRK and QMRK subprocesses to give equal contributions to $B$-meson production.

\begin{figure}[ht]
\begin{center}
\includegraphics[width=1.0\textwidth]{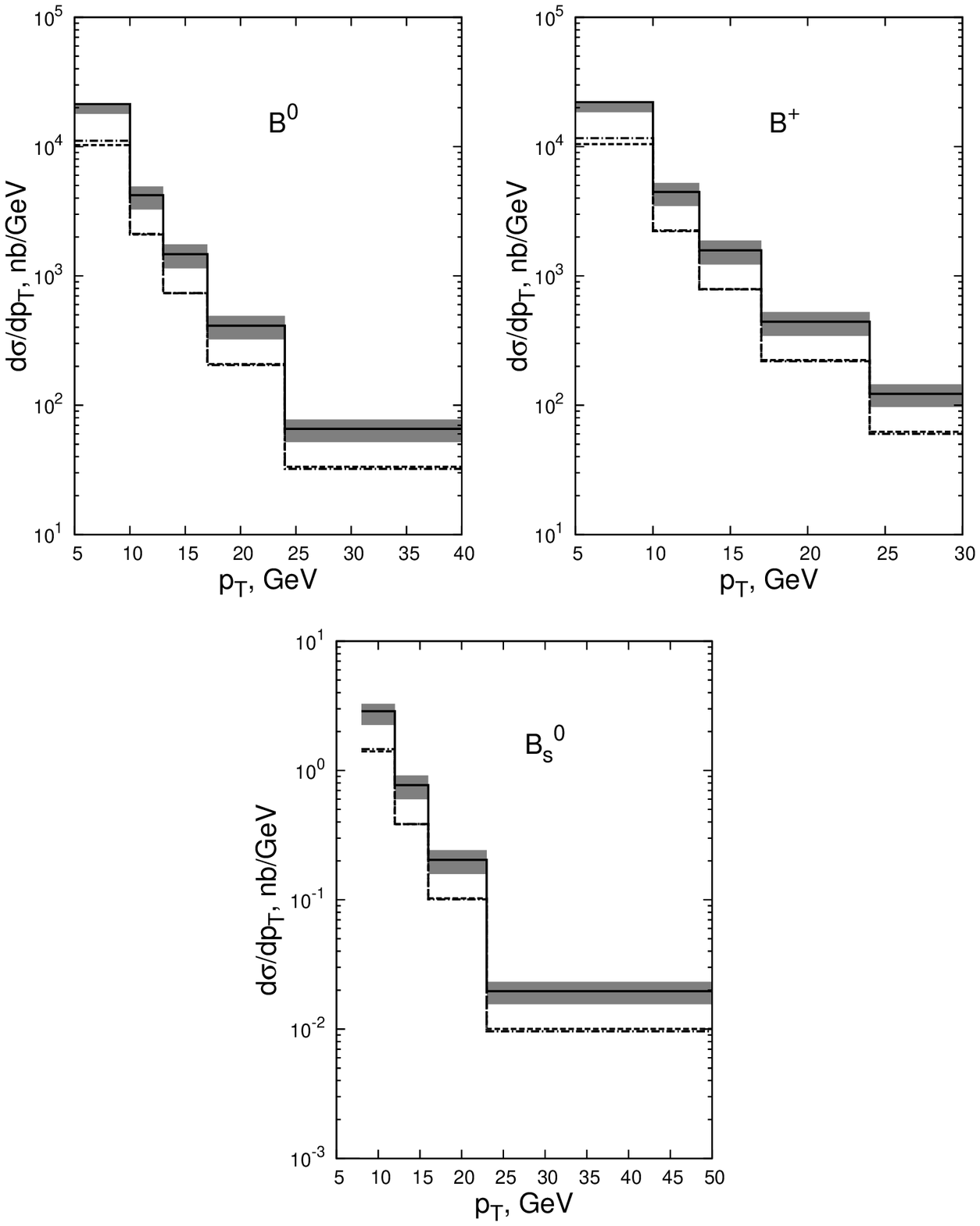}
\end{center}
\caption{Theoretical predictions for the transverse momentum distributions of $B^0$ (top), $B^+$ (middle), $B^s$ (bottom) mesons
in $pp$ scattering at $\sqrt S=14$~TeV and $|y|< 1.0$ obtained in the LO PRA. The notations as in the Fig.~\ref{fig:central}.
 \label{fig:14}}
\end{figure}

Finally, in the Fig.~\ref{fig:14}
we present our predictions for the planned LHC energy of $\sqrt
S=14$~TeV and keeping the other kinematic conditions as in the
Refs.~\cite{CMSB0,CMSBplus,CMSBs}.

\begin{figure}[ph]
\begin{center}
\includegraphics[width=1.0\textwidth]{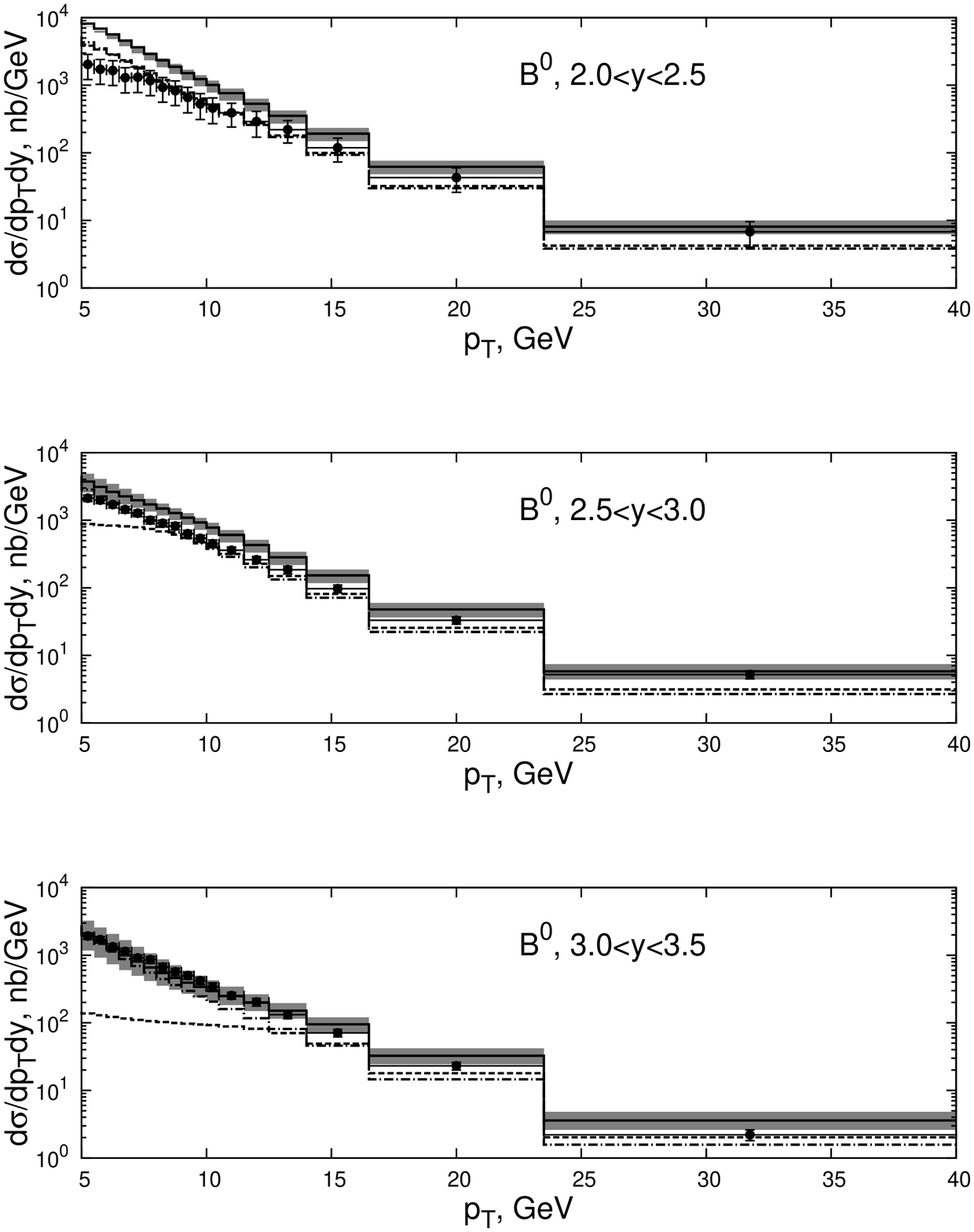}
\caption{ Transverse momentum distributions of $B^0$ mesons in the forward rapidity regions of $2.0<y<2.5$ (top), $2.5<y<3.0$ (middle), and $3.0<y<3.5$ (bottom)
in $pp$ scattering with $\sqrt S=7$ TeV. The LHCb data at LHC are from the Ref.~\cite{LHCb}. The notations as in the Fig.~\ref{fig:central}.\label{fig:BRB0}}
\end{center}
\end{figure}

\begin{figure}[ph]
\begin{center}
\includegraphics[width=1.0\textwidth]{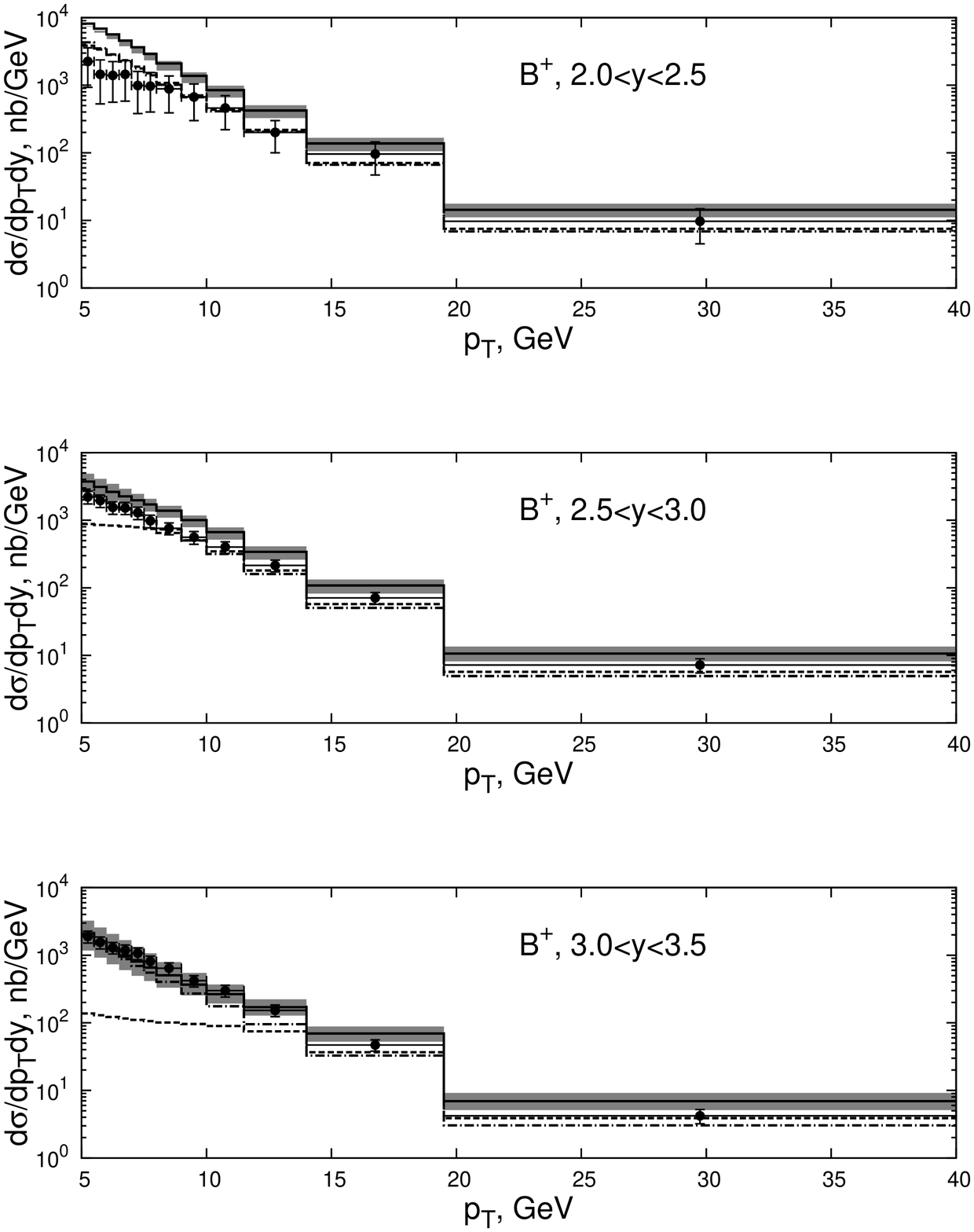}
\caption{ The same as in the Fig.~\ref{fig:BRB0} for $B^+$ mesons}.\label{fig:BRBplus}
\end{center}
\end{figure}

\begin{figure}[ph]
\begin{center}
\includegraphics[width=1.0\textwidth]{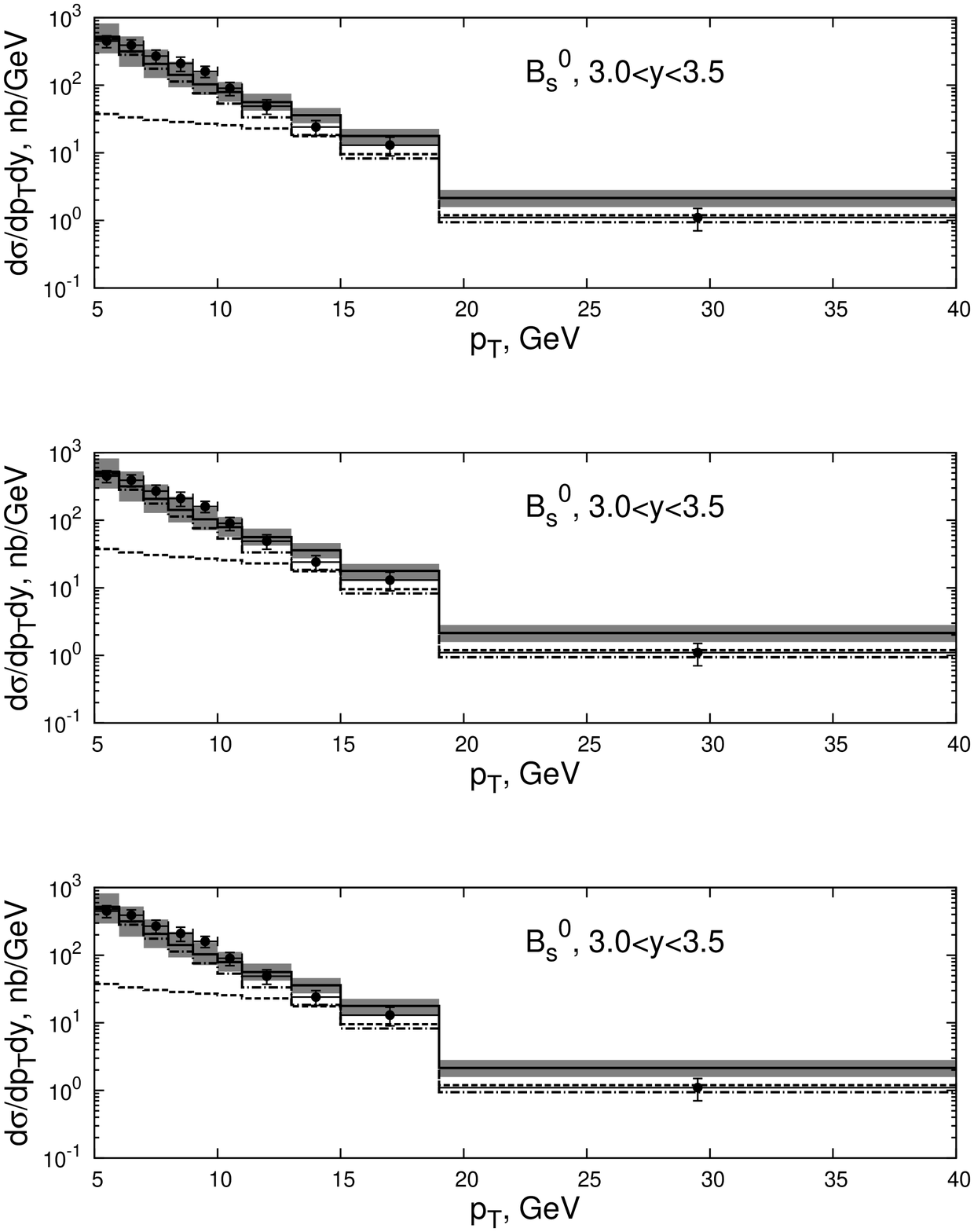}
\caption{ The same as in the Fig.~\ref{fig:BRB0} for $B^0_s$ mesons}.\label{fig:BRs1}
\end{center}
\end{figure}

\begin{figure}[ph]
\begin{center}
\includegraphics[width=1.0\textwidth]{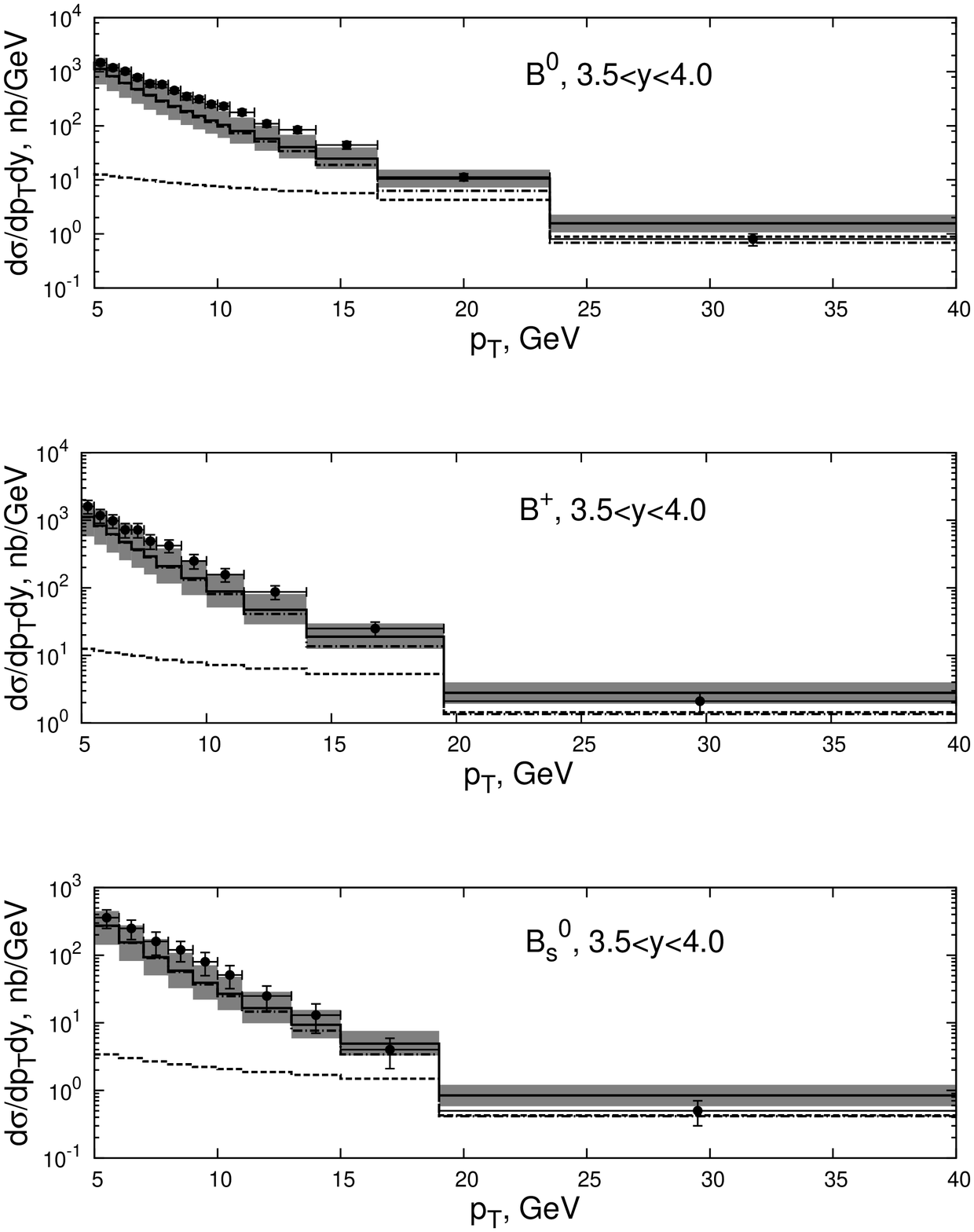}
\caption{ Transverse momentum distributions of $B^0$, $B^+$, and $B_s^0$ mesons in the forward rapidity region of $3.5<y<4.0$
in $pp$ scattering with $\sqrt S=7$ TeV. The LHCb data at LHC are from the Ref.~\cite{LHCb}. The notations as in the Fig.~\ref{fig:central}. \label{fig:BRB4}}
\end{center}
\end{figure}

\begin{figure}[ph]
\begin{center}
\includegraphics[width=1.0\textwidth]{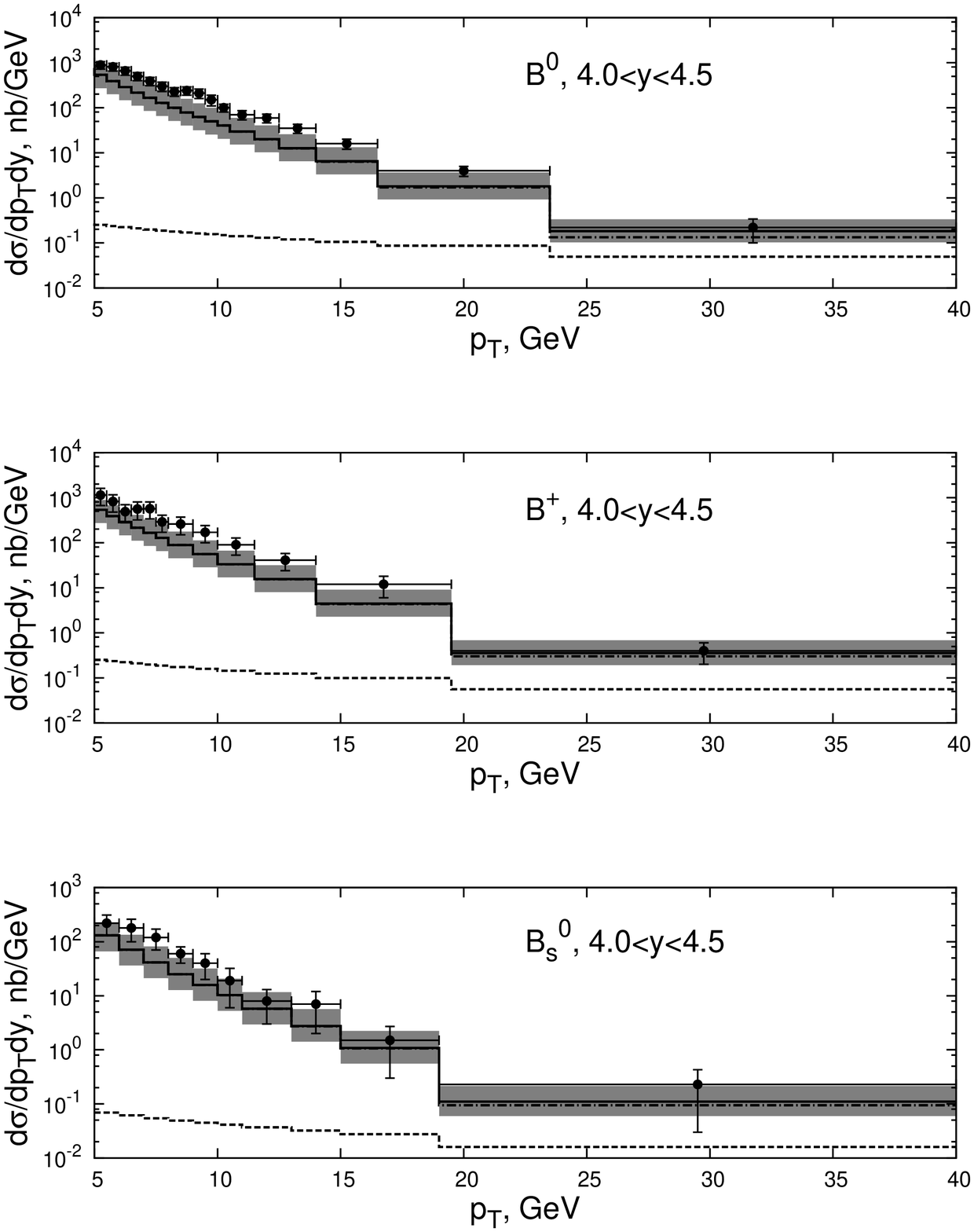}
\caption{ The same as in the Fig.~\ref{fig:BRB4} for $4.0<y<4.5$.\label{fig:BRB5}}
\end{center}
\end{figure}

\begin{figure}[ph]
\begin{center}
\includegraphics[width=1.0\textwidth]{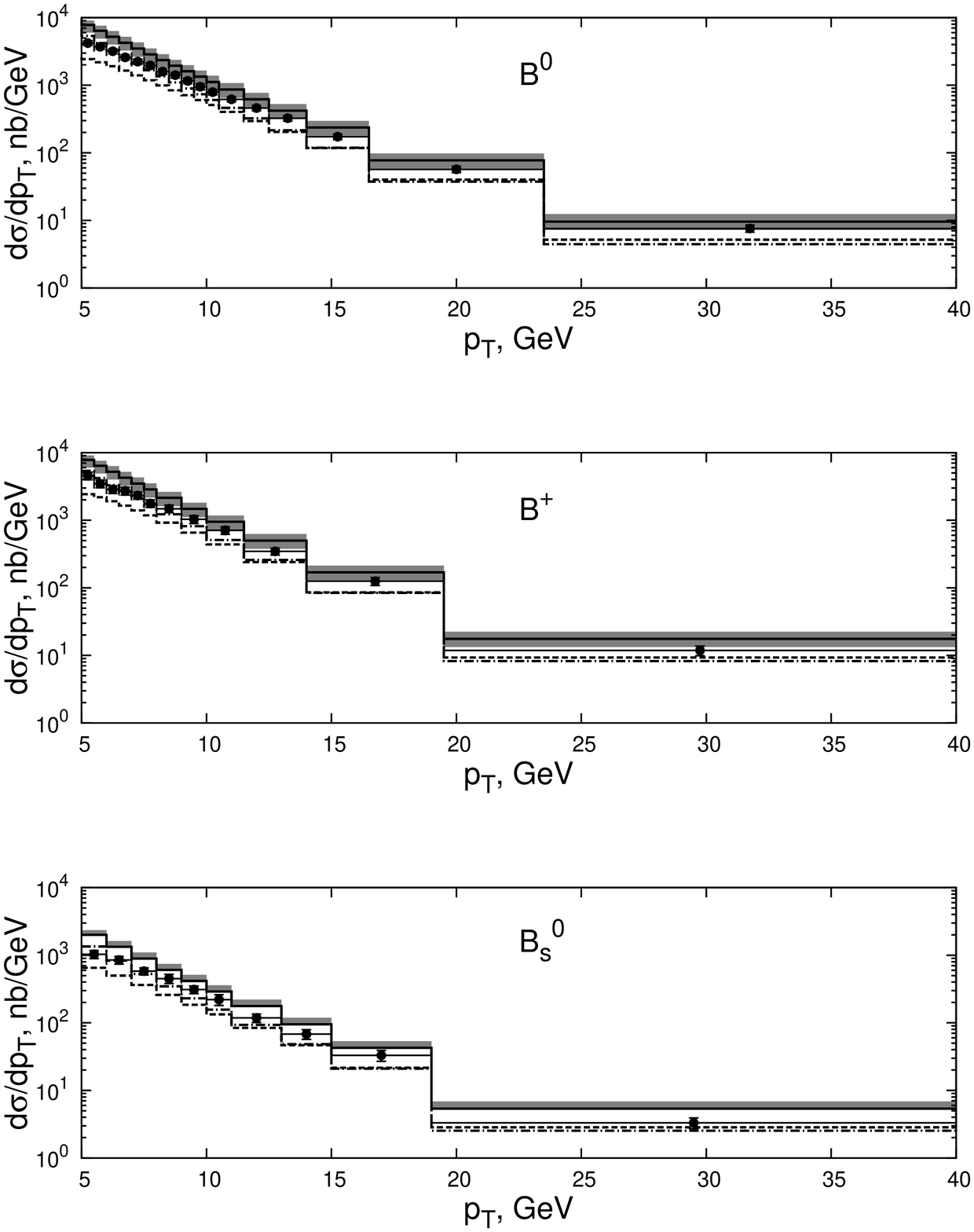}
\caption{ Transverse momentum distributions of $B^0$ (left-top), $B^+$ (right-top), $B^0_s$ (bottom) mesons in the forward rapidity region
in $pp$ scattering with $\sqrt S=7$~TeV and $2.0<y<4.5$. The LHCb data at LHC are from the Ref.~\cite{LHCb}. The notations as in the Fig.~\ref{fig:central}.\label{fig:BR0}}
\end{center}
\end{figure}

Not only the central but also the forward rapidity region in $pp$
collisions at the LHC became available by the specially designed
LHCb detector where the measurements of differential cross sections
of $B^0$, $B^+$, and $B_s^0$ mesons including their charge-conjugate states were performed
at $\sqrt S=7$~TeV with $2.0<y<4.5$~\cite{LHCb}. The observed data
divided into 5 rapidity regions previously found a successful description in the FONLL
scheme~\cite{newNLO}. We present these
data coming as double-differential distribution for the each of rapidity regions
and the transverse momenta distribution integrated over all considered rapidities,
together with our results obtained in the LO of PRA in the
Figs.~\ref{fig:BRB0}-\ref{fig:BR0}. Nevertheless the PRA formalism
discussed here is justified for the particle production in the central interval of rapidities,
we obtain a good description of the data at $p_T\geq 10$~GeV
even in the forward rapidity region.
As for the region of the transverse momenta comparable to a $b$-quark mass,
at the lower forward rapidities our predictions overestimate the experimental data.
Moving towards the higher rapidities this excess transforms to the underestimation
due to the strong decreasing of the contribution of subprocess~(\ref{eq:RRg}) at small $p_T$
with increasing rapidity. This behaviour of gluon contribution differs the overall picture
at forward rapidities from the central one, where the contributions of both subprocesses are approximately equal.

This effect can  be explained if we
we recall that with grow of
rapidity of the particle produced in the hard scattering process the
fraction of longitudinal momenta of initial proton transferred to
this process increases simultaneously. That means
that we enter the region of large $x>0.1$ where the conditions of
Reggeization are not satisfied and the CPM should be applied instead of PRA.
The large positive rapidity of a produced particle is provided by a large fraction $x$
from the hadron moving along the positive direction, and balanced by a very
small fraction of negative longitudinal momenta carrying by the second hadron in the collision, up to $10^{-5}$.
That leads to the situation in which we
finely take into account small-$x$ effects although loosing in
large-$x$. The region of large $x$ and small $p_T$ is a field of study for CPM,
where the $2\to 1$ processes with non-zero transverse momentum of final particle do not exist
and one should start from $2\to 2$ processes. Such a way, the contribution of the subprocess~(\ref{eq:RRg}) falls down, and
the subprocess~(\ref{eq:RRQQ}) starts to give a dominant contribution,
while the underestimating of the data is connected with the NLO corrections to the latter which are beyond this study.
Moreover, as one can find from Fig.~\ref{fig:FFB}, the $b$-quark fragmentation function strongly exceeds the gluon one, especially at low $\mu^2$. This fact is confirmed by our recent work~\cite{bbLHC}, where the bottom quark multiplicity in a gluon jet for different $\mu^2$ was extracted.


\section{Conclusions}
\label{sec:four}

In the present work we performed the study of  $B^0$,
$B^+$, and $B_s^0$-meson fragmentation production in proton-(anti)proton collisions with central rapidities
at Tevatron Collider and LHC, and in the forward rapidity region for the LHC, in the framework of Parton Reggeization Approach. Here we take into account all the hard-scattering parton subprocesses appearing at the LO with Reggeized gluons in the initial state. Among them there is a $2\to 1$ subprocess of gluon production via Reggeized-gluon fusion, which was considered at the first time during the studies of $B$-meson production. To describe the hard scattering stage we use the fully gauge invariant amplitudes introduced in the works of L.~N.~Lipatov and co-authors. The distributions of initial partons are taken in the form of unintegrated parton distribution functions proposed by Kimber, Martin and Ryskin, and the way of their definition is ideologically related to the above-mentioned amplitudes.
 To describe the non-perturbative transition of gluons and $b$-quarks created at the hard stage into the $B$-mesons we use the universal fragmentation functions obtained from the fit
  of $e^+ e^-$ annihilation data from CERN LEP1 and SLAC SLC colliders.
 We obtained a good agreement of our results for $B$-meson central-rapidity production comparing with experimental data from Tevatron and the LHC,
  especially at large transverse momenta. The achieved degree of agreement for the central rapidity region is the same as the one obtained
   by NLO calculations in
  the conventional collinear parton model. The predictions for the
  $B$-meson production at central rapidities for the expected LHC energy of $\sqrt S=14$~TeV are also presented.
   At the forward rapidities our results for the transverse-momentum $B$-meson distributions are found to diverge with experimental data provided by the LHCb Collaboration at LHC, that effect is . We describe
  $B$-meson production without any free parameters or auxiliary approximations.

\section{Acknowledgements}
\label{sec:five} The work of A.~V.~Shipilova and A.~V.~Karpishkov
was partly supported by the Grant of President of Russian Federation
No. MK-4150.2014.2. The work of M.A.~Nefedov and V.A.~Saleev  was
supported in part by the Russian Foundation for Basic Research
through the Grant No. 14-02-00021. A.~V.~Shipilova is grateful
to Prof. G.~Kramer for the useful discussions, to Prof. B.~A.~Kniehl for the kind hospitality,
and to the German Academic Exchange Service (DAAD) together with the Russian Federal Ministry of Science and Education
for the financial support by Grant~No.~A/13/75500.

\newpage

\newpage

\newpage

\newpage

\newpage

\newpage

\newpage

\newpage

\newpage

\newpage

\end{document}